\documentclass[twocolumn,showpacs,aps,prl,superscriptaddress,letterpaper,floatfix]{revtex4}
\usepackage{graphicx}
\usepackage{dcolumn}
\usepackage{amsmath}
\usepackage{epsfig}

\newcommand{\BABARPubYear}       {07}
\newcommand{\BABARPubNumber}     {013}
\newcommand{\SLACPubNumber} {12387}

\input babarsym.tex

\def\pipi     {\ensuremath{\pi\pi}}

\def\akpi  {\ensuremath{{\cal A}_{K\pi}}}
\def\akpiraw  {\ensuremath{{\cal A}_{K\pi}^{\rm raw}}}

\def\fpm {\ensuremath{f_{\pm}(\deltat)}}

\def\spipi {\ensuremath{S_{\pi\pi}}}
\def\cpipi {\ensuremath{C_{\pi\pi}}}
\def\de {\ensuremath{\Delta E}}

\def\Btag {\ensuremath{B_{\rm tag}}}

\def\Bflav {\ensuremath{B_{\rm flav}}}

\RequirePackage{xspace}

\def\figurebox#1#2#3{%
    \def\arg{#3}%
    \ifx\arg\empty
    {\hfill\vbox{\hsize#2\hrule\hbox to #2{\vrule\hfill\vbox to #1{\hsize#2\vfill}\vrule}\hrule}\hfill}%
    \else
    {\hfill\epsfbox{#3}\hfill}%
    \fi}

\long\def\inst#1{\par\nobreak\kern 4pt\nobreak
    {\it #1}\par\vskip 10pt plus 3pt minus 3pt}

\begin{document}

\begin{flushleft}
\babar-PUB-\BABARPubYear/\BABARPubNumber\\
SLAC-PUB-\SLACPubNumber\\
\end{flushleft}

\title{
{
\Large \bf \boldmath
 Observation of \CP Violation in $\Bz\to\Kp\pim$ and $\Bz\to\pip\pim$ 
}}

%
\author{B.~Aubert}
\author{M.~Bona}
\author{D.~Boutigny}
\author{Y.~Karyotakis}
\author{J.~P.~Lees}
\author{V.~Poireau}
\author{X.~Prudent}
\author{V.~Tisserand}
\author{A.~Zghiche}
\affiliation{Laboratoire de Physique des Particules, IN2P3/CNRS et Universit\'e de Savoie, F-74941 Annecy-Le-Vieux, France }
\author{J.~Garra~Tico}
\author{E.~Grauges}
\affiliation{Universitat de Barcelona, Facultat de Fisica, Departament ECM, E-08028 Barcelona, Spain }
% ...
\collaboration{The \babar\ Collaboration}
\noaffiliation

\date{8 March 2007; published 12 July 2007}

\begin{abstract}
We report observations of \CP violation in the decays 
$\Bz\to\Kp\pim$ and $\Bz\to\pip\pim$ in a sample of
383 million $\Y4S\to\BB$ events. We find 
$4372\pm 82$ $\Bz\to \Kp\pim$ decays and 
measure the direct \CP-violating charge asymmetry
$\akpi =  -0.107 \pm 0.018\,({\rm stat}) ^{+0.007}_{-0.004} \,({\rm syst})$,
which excludes the \CP-conserving hypothesis 
with a significance of 5.5 standard deviations. 
In the same sample we find $1139 \pm 49$ $\Bz\to\pip\pim$ decays and measure 
the \CP-violating asymmetries 
$\spipi = -0.60\pm 0.11\,({\rm stat})\pm 0.03\,({\rm syst})$ and 
$\cpipi = -0.21\pm 0.09\,({\rm stat})\pm 0.02\,({\rm syst})$. 
\CP conservation in $\Bz\to\pip\pim$ $(\spipi=\cpipi=0)$ is excluded 
at a confidence level $1-\textrm{C.L.} = 8 \times 10^{-8}$, corresponding 
to 5.4 standard deviations.
\end{abstract}


\maketitle


The prediction of large \CP-violating effects in the $B$-meson
system~\cite{largeCPV} has been confirmed in recent years by the \babar\
and Belle collaborations, both in the interference of $B$ decays to
charmonium final states with and without $\Bz$--$\Bzb$
mixing~\cite{CPV_beta}, and directly in the interference between the
decay amplitudes in $\Bz\to\Kp\pim$~\cite{BaBarAkpiPRL,BelleKpiRef,refConj}.  
All measurements of \CP\ violation to date are in agreement with indirect
predictions from global standard-model (SM) fits~\cite{SMfits} based on
measurements of the magnitudes of the elements of the
Cabibbo--Kobayashi--Maskawa (CKM) quark-mixing matrix~\cite{CKM}
and place important constraints~\cite{CKMnewphys} on the flavor
structure of SM extensions.

The proper-time evolution of the asymmetry between \Bz and \Bzb decays to $\pip\pim$
is characterized by sine and cosine terms with amplitudes \spipi, which arises from 
interference between decays with or without \Bz--\Bzb mixing, and \cpipi,
which is due to interference between the $b \to u$ ``tree'' and the higher-order $b \to d$ 
``penguin'' decay amplitudes.  
Similarly, the direct-\CP-violating asymmetry 
\akpi\ between 
the $\Bzb\to\Km\pip$ and $\Bz\to\Kp\pim$ decay rates arises from interference 
between $b \to u$ tree and $b \to s$ penguin amplitudes.  
Negligible contributions to these asymmetry parameters would also enter from 
\CP violation purely in \Bz--\Bzb mixing, which has been determined to be very
small~\cite{PDG2006}. 
The quantity 
$\sin{2\alpha_{\rm eff}} = \spipi/\sqrt{1-\cpipi^2}$ can be related to 
$\alpha \equiv \arg\left[-V_{\rm td}^{}V_{\rm tb}^{*}/V_{\rm ud}^{}V_{\rm ub}^{*}\right]$
through a model-independent analysis that uses the isospin-related decays 
$B^{\pm}\to\pipm\piz$ and $\Bz\to\piz\piz$~\cite{alphafrompenguins}.
Contributions from new particles could affect the asymmetries in these modes
primarily through additional penguin $B$-decay amplitudes. 

Previous evidence of direct \CP\ violation in $\Bz\to\Kp\pim$ has been reported by 
\babar~\cite{BaBarAkpiPRL} and Belle~\cite{BelleKpiRef}; additional measurements 
of $\akpi$ have also been reported by the CDF~\cite{CDFKpiRef} and CLEO~\cite{CLEOKpiRef} 
Collaborations.  The Belle Collaboration recently reported~\cite{BellePiPi2006} an 
observation of both time-dependent and direct \CP violation in $\Bz\to\pip\pim$ decays
using a sample of $535 \times 10^6$ \BB pairs, while our previous measurement~\cite{BaBarPiPi2004}
on a sample of $227 \times 10^6$ \BB pairs was statistically consistent with no \CP violation. 
In this Letter, we present measurements of \akpi, \spipi, and \cpipi\ in a sample of
$ 383 \times 10^6$ \BB pairs using an improved analysis technique with significantly
increased sensitivity compared to our previous measurements.

In the \babar\ detector~\cite{ref:babar}, charged particles are detected 
and their momenta measured by a combination of a five-layer silicon vertex tracker (SVT)
and a 40-layer drift chamber (DCH) that covers 92\% of the solid angle
in the \Y4S center-of-mass (c.m.) frame, both operating in a 1.5-T solenoidal magnetic field.
Discrimination between charged pions, kaons, and protons is provided by a combination of an 
internally reflecting ring-imaging Cherenkov detector (DIRC), which covers 84\% of the c.m. solid 
angle in the central region of the \babar\ detector and has a 91\% reconstruction efficiency 
for pions and kaons with momenta above 1.5 \gevc, and the 
ionization (\dedx) measurements in the DCH. Electrons 
are explicitly removed based on a comparison of the track momentum and the associated energy 
deposition in a CsI(Tl) electromagnetic calorimeter, and with additional 
information from \dedx\ and DIRC Cherenkov angle $(\theta_{\rm C})$ measurements. 

The analysis method retains many features of our previous $\Bz\to\Kp\pim$ and $\Bz\to\pip\pim$
\CP-violation measurements~\cite{BaBarAkpiPRL,BaBarPiPi2004}. We reconstruct candidate decays
$B_{\rm rec} \to h^+ h^-$ $(h^{\pm} = \pipm,\,\,\Kpm)$ from pairs of oppositely charged tracks 
in the polar-angle range $0.35 < \theta_{\rm lab} <2.40 $ that are consistent with originating 
from a common decay point.
The remaining particles are examined to infer (flavor tag) whether 
the other $B$ meson in the event (\Btag) decayed as a $\Bz$ or $\Bzb$.
We perform an unbinned extended maximum-likelihood (M.L.) fit simultaneously for the 
\CP-violating asymmetries and the signal and background yields and parameters.  
The fit uses particle-identification, kinematic, event-shape, $B_{\rm tag}$ flavor, and 
$\deltat$ information, where $\deltat$ is the difference between the $B_{\rm rec}$ and 
$B_{\rm tag}$ decay times.  The yields for the $K\pi$ final state are parametrized as 
$n_{\Kpm\pimp}=n_{K\pi}\left(1\mp \akpiraw\right)/2$, and the 
decay-rate distribution $f_+\,(f_-)$ for $B_{\rm rec}\to\pip\pim$ and 
$\Btag = \Bz\,(\Bzb)$ is given by 
\begin{eqnarray}
\fpm = \frac{e^{-\left|\deltat\right|/\tau}}{4\tau} [1
& \pm & \spipi\sin(\deltamd\deltat) \nonumber \\
& \mp & \cpipi\cos(\deltamd\deltat)],
\label{fplusminus}
\end{eqnarray}
where $\tau$ is the neutral $B$\/ lifetime and $\deltamd$ is the \Bz--\Bzb mixing
frequency, both fixed to their world averages~\cite{PDG2006}.

The most significant improvement in sensitivity compared to our previous analysis
comes from a 35\% increase in the $B_{\rm rec}$ reconstruction efficiency that 
results from using \dedx\ as a discriminating variable in the 
M.L.\ fit for the first time.  The \dedx\ measurements are used both to 
complement the discriminating power of $\theta_{\rm C}$ for charged particles within 
the DIRC acceptance and as a standalone means of particle identification for tracks that have 
no DIRC information and were not included in our previous measurements.  
The \dedx\ calibration takes into account variations in the mean value and resolution 
of \dedx\ with respect to changes in the DCH running conditions over time and each 
track's charge, polar and azimuthal angles, and number of ionization samples.  The 
calibration is performed with large ($> 10^6$) high-purity samples of protons from 
$\Lambda\to\proton\pim$, pions and kaons from $D^{*+}\to D^0\pi^+\,(\Dz\to\Km\pip)$, 
and additional samples of pions from $\tau^{-} \to \pim\pip\pim\nu_{\tau}$ decays and from 
$\KS \to\pip\pim$ decays that occur in the vicinity of the interaction region. 

\begin{figure}[!b]
\includegraphics[width=0.95\linewidth,clip=true]{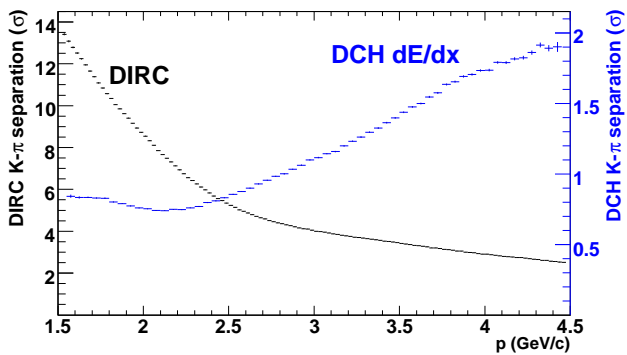}
\caption{\small The average difference between the expected values of DIRC $\theta_{\rm C}$ 
and DCH \dedx\ 
for pions and kaons at $0.35 < \theta_{\rm lab} <2.40 $, divided by the uncertainty, 
as a function of laboratory momentum in $\Bz\to h^+h^-$ decays in $\babar$.}
\label{fig1}
\end{figure}

We require at least one of the tracks in the $B_{\rm rec}$ decay candidate 
to have $\theta_{\rm C}$ measured with at least six signal photons; for such tracks, 
the value of $\theta_{\rm C}$ must agree within 4 standard deviations ($\sigma$) with 
either the pion or kaon hypothesis.  
Thus, protons with six or more signal photons
are removed while proton-pion and proton-kaon combinations are possible 
for background candidates where one of the tracks has no usable $\theta_{\rm C}$ measurement. 
We construct $\theta_{\rm C}$ probability-density functions (PDFs)
for the pion and kaon hypotheses, and \dedx\ PDFs for the 
pion, kaon, and proton hypotheses, separately for each charge. 
The $K$--$\pi$ separations provided by $\theta_{\rm C}$ and \dedx\ are complementary:  
for $\theta_{\rm C}$, it varies from $2.5 \sigma$ at $4.5\gevc$ to $13 \sigma$ at 
$1.5 \gevc$~\cite{BaBarAkpiPRL}, while for \dedx\ it varies from less than 
$1.0\sigma$ at $1.5\gevc$ to $1.9 \sigma$ at $4.5 \gevc$ (Fig.~\ref{fig1}).

Each $B$\/ candidate is characterized by the energy 
difference $\Delta E = (q_{\Upsilon} \cdot q_{B}/\sqrt{s}) - \sqrt{s}/2$, 
which also provides additional discriminating power between the 
four possible final states ($\pip\pim$, $\Kp\pim$, $\Km\pip$, $\Kp\Km$), and
the beam-energy--substituted mass
$\mes = [({s}/{2} + {\vec{p}}_{\Upsilon} \cdot 
{\vec{p}}_{B} )^{2}/{E^{2}_{\Upsilon}} - 
{{\vec{p}}_{B}}^{\,2}]^{1/2}$~\cite{ref:babar}.
Here, $q_{\Upsilon}$\/ and $q_{B}$ are the four-momenta of the \Y4S and the $B$\/ 
candidate, $s \equiv (q_{\Upsilon})^{2}$ is the square of the 
c.m. energy, ${\vec{p}}_{\Upsilon}$ and ${\vec{p}}_{B}$ are the laboratory three-momenta 
of the \Y4S\/ and the $B$, 
and ${E_{\Upsilon}} \equiv q^0_{\Upsilon}$ is the laboratory energy of the \Y4S.
For signal events, the \mes and \DeltaE PDFs are Gaussian functions with 
widths of $2.6\mevcc$ and $29\mev$, respectively.
For the background, \mes is parametrized with an empirical 
threshold function~\cite{ref:argus} and \DeltaE is parametrized
with a second-order polynomial.
We require $5.2<\mes<5.3$ \gevcc and $\left|\de\right| < 0.150$ \gev.

The background arises predominantly from random combinations of tracks 
in $\epem\to q\bar{q}$ $(q=u,d,s,c)$ and $\tau^+\tau^-$ jetlike continuum events. 
We define the angle $\theta_S$ in the c.m. frame between the sphericity axes~\cite{sph} of 
the $B$\/ candidate and of all remaining charged and neutral particles in the event. For 
background events, $\left|\cos{\theta_S}\right|$ peaks sharply near 1, while for 
$B$\/ decays the distribution is nearly flat.  We require $\left|\cos{\theta_S}\right|<0.9$, 
which removes approximately 64\% of \uubar, \ddbar, and \ssbar, 52\% of \ccbar, and 84\% 
of $\tau^+\tau^-$ background.  
Contamination from $\epem\to \tau^+\tau^-$ production is reduced to 2\% of the total background
by requiring the ratio of the second to
zeroth Fox--Wolfram moments~\cite{R2all} to be less than 0.7,
which has a negligible effect on the signal efficiency.
The overall gain in signal reconstruction efficiency is 52\% compared to our previous analysis.
Additional continuum-background suppression in the 
fit is accomplished by the Fisher discriminant ${\cal F}$ described in Ref.~\cite{BaBarSin2alpha2002}.
We have studied the backgrounds from higher-multiplicity $B$ decays 
and find them to be negligible, particularly due to their good separation from signal in \DeltaE.  

The $\Btag$ flavor is determined with a neural-net algorithm~\cite{BaBarsin2beta} that
assigns the event to one of seven mutually exclusive tagging categories.
The figure of merit for the tagging quality, measured in a data sample \Bflav\ 
of fully reconstructed \Bz decays to $D^{(*)-}(\pip,\, \rho^+,\, a_1^+)$ or
$\jpsi\Kstarz$, 
is the effective efficiency 
$Q = \sum_k \epsilon_k (1-2w_k)^2 = 0.305 \pm 0.003$, where $\epsilon_k$ and $w_k$ are the 
efficiencies and mistag probabilities for events in tagging category $k$.
Separate values of $\epsilon_k$ and $w_k$ for each background category 
are determined in the M.L.\ fit.

The time difference $\deltat \equiv \Delta z/\beta\gamma c$, where $\beta\gamma \approx 0.56$
is the known boost of the \Y4S, is obtained by measuring the distance 
$\Delta z$ along the beam ($z$) axis between the $B_{\rm rec}$ and $B_{\rm tag}$ decay
vertices.  We require $\left|\deltat\right|<20\ps$ and 
$\sigma_{\deltat} < 2.5\ps$, where $\sigma_{\deltat}$ is the $\deltat$ uncertainty estimated
separately for each event.  The resolution function for signal candidates is a sum of 
three Gaussians~\cite{BaBarsin2beta} with parameters determined from a fit to the full \Bflav\ sample.
The background $\deltat$ distribution, common to all tagging categories, is modeled
as a sum of three Gaussian functions with parameters determined in the final fit. 

The likelihood for candidate $j$ tagged in category 
$k$ is obtained by summing the product of event yield $n_{i}$, tagging efficiency 
$\epsilon_{i,k}$, and probability ${\cal P}_{i,k}$ over all possible signal 
and background hypotheses $i$.  We treat separately the cases where both or only one track has a 
$\theta_{\rm C}$ measurement.  The extended likelihood function 
for tagging category $k$ is
\begin{equation}
{\cal L}_k = \exp{\left(-\sum_{i}n_i\epsilon_{i,k}\right)}
\prod_{j}\left[\sum_{i}n_i\epsilon_{i,k}{\cal P}_{i,k}(\vec{x}_j;\vec{\alpha}_i)\right].
\end{equation}
The probabilities ${\cal P}_{i,k}$ are evaluated as a product of PDFs 
for each of the independent variables 
$\vec{x}_j = \left\{\mes, \de, {\cal F}, \dedx, \theta_{\rm C}, \deltat\right\}$, 
with parameters $\vec{\alpha}_i$.  We use separate $\theta_{\rm C}$ and \dedx\ PDFs for
positively- and negatively-charged tracks.  The $\deltat$ PDF for signal $\pip\pim$ decays is given 
by Eq.~(\ref{fplusminus}) modified to include the mistag probabilities for each tagging 
category and convolved with the signal resolution function.  The $\deltat$ PDFs for 
signal $K\pi$ and background $K\pi$, $\pi\proton$, and $K\proton$
combinations take into account the correlation 
between the charge of the kaon or proton and the $\B_{\rm tag}$ flavor; 
for signal $K\pi$, \Bz--\Bzb mixing is also taken into account.
The total likelihood 
${\cal L}$ is the product of likelihoods for each tagging category and has 117 free parameters.

\begin{figure}[!b]
\includegraphics[width=0.75\linewidth,clip=true]{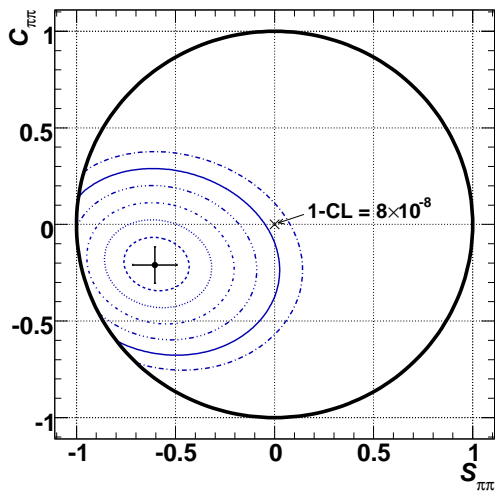}
\caption{\small \spipi\ and \cpipi: the central values, errors, and confidence-level (C.L.) contours for 
$1-\mathrm{C.L.} = 0.317$ $(1\sigma)$, $4.55 \times 10^{-2}$ $(2\sigma)$, $2.70 \times 10^{-3}$ $(3\sigma)$, 
$6.33 \times 10^{-5}$ $(4\sigma)$, $5.73 \times 10^{-7}$ $(5\sigma)$, and $1.97 \times 10^{-9}$ $(6\sigma)$,
calculated from the square root of the change in the value of
$-2\ln{\cal L}$ compared with its value at the minimum.
The systematic errors are included.}
\label{fig2}
\end{figure}

Fitting the final sample of $309540$ events, 
we find $n_{\pi\pi} = 1139\pm 49$, $n_{K\pi} = 4372\pm 82$, $n_{KK}=10\pm 17$, 
where all errors are statistical only, and measure the following asymmetries:
\begin{eqnarray*}
\akpi & =          & -0.107\pm 0.018\,({\rm stat})^{+0.007}_{-0.004}\,({\rm syst}),\\
\spipi & =          & -0.60\pm 0.11\,({\rm stat})\pm 0.03\,({\rm syst}),\\
\cpipi & =          & -0.21\pm 0.09\,({\rm stat})\pm 0.02\,({\rm syst}).
\end{eqnarray*}
Here \akpi\ is the fitted value of the $\Kmp\pipm$ event-yield asymmetry \akpiraw\ 
shifted by $+0.005^{+0.006}_{-0.003}$ to account for a bias that arises 
from the difference between the cross sections of \Kp and \Km hadronic interactions
within the \babar\ detector.  We determine this bias from a detailed Monte Carlo 
simulation based on GEANT4~\cite{GEANT4} version 7.1; it is independently verified with
a calculation based on the known material composition of the 
\babar\ detector~\cite{ref:babar} and the cross sections and material properties 
tabulated in Ref.~\cite{PDG2006}. The corrected $\Kmp\pipm$ event-yield asymmetry in the 
background, where no observable \CP\ violation is expected, is consistent with zero:
$-0.006 \pm 0.004\, (\rm stat) ^{+0.006}_{-0.003}\, (\rm syst)$.

\begin{figure}[!b]
\includegraphics[width=0.75\linewidth,clip=true]{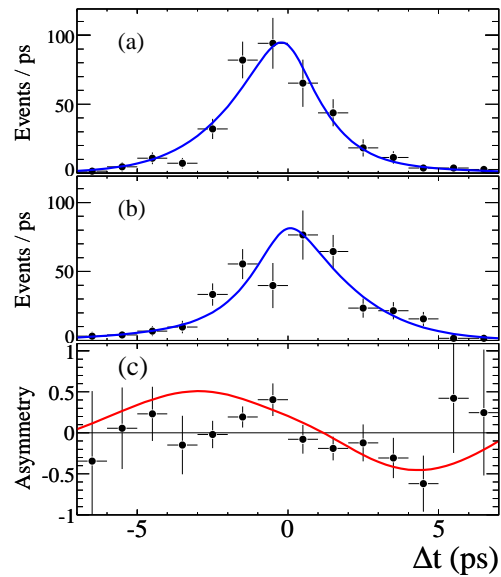}
\caption{\small The background-subtracted distributions of the decay-time 
difference $\deltat$ in signal $B\to\pip\pim$ events.
The points with errors show the events where $B_{\rm tag}$ is
identified as (a) $\Bz$ or (b) $\Bzb$. The asymmetry, defined as 
$\left(n_{\Bz} - n_{\Bzb}\right)/\left(n_{\Bz} + n_{\Bzb}\right)$, for signal
events in each $\deltat$ bin, is shown in (c). 
The solid curves are the projection of the fit.}
\label{fig:asym}
\end{figure}

A contour plot of the $(\spipi,\cpipi)$ confidence levels is shown 
in Fig.~\ref{fig2}.  The correlation between \spipi\ and \cpipi\ is $-0.07$. 
Performing a fit that excludes $\deltat$ and using an 
event-weighting technique~\cite{sPlots}, in Fig.~\ref{fig:asym} we show the distributions of 
$\deltat$ for signal $\pip\pim$ events with $B_{\rm tag}$ tagged as $\Bz$ or $\Bzb$, 
and the asymmetry as a function of $\deltat$, overlaid with the PDF curves 
that represent the result of the full fit.

To validate our results, we perform a number of consistency checks and 
systematic-error studies similar to those reported in Refs.~\cite{BaBarAkpiPRL} 
and~\cite{BaBarPiPi2004}.  For \akpi, the dominant source of systematic 
uncertainty is the bias due to kaon hadronic interactions.
We find that the systematic errors due to potentially imperfect understanding 
of the DIRC and DCH particle-identification performance
are small for \akpi\ $(0.002)$, \spipi\ $(0.007)$, and \cpipi\ $(0.006)$.  
The dominant sources of systematic uncertainty on $\spipi$ are the signal $\deltat$
model ($0.020$) and flavor-tagging parameters ($0.015$), while for $\cpipi$ the
dominant uncertainties arise from tagging ($0.014$) and the potential effect~\cite{Owen} 
of doubly CKM-suppressed decays of the $B_{\rm tag}$ meson $(0.016)$.
As a final cross-check, we perform a fit allowing the mixing frequency and
lifetime to vary simultaneously with $\spipi$ and $\cpipi$.  We find
$\deltamd=0.506\pm 0.017\,{\rm ps}^{-1}$ and 
$\tau_{\Bz}=1.523\pm 0.026\,{\rm ps}$,
where the errors are statistical only, 
consistent with the world-average values, and the resulting shifts in 
the \CP\ parameters are negligible.  The total systematic uncertainties are 
calculated by summing all individual contributions in quadrature.

In summary, we observe direct \CP violation in the decay $\Bz\to\Kp\pim$ 
with a statistical significance of $5.5 \sigma$ and \CP\ violation in the 
time distribution of $\Bz\to\pip\pim$ decays with a significance 
of $5.4 \sigma$. We also determine that the mixing-induced \CP-violating 
asymmetry \spipi\ is nonzero with a significance 
of $5.1 \sigma$ or greater for any value of \cpipi. 
All results are consistent with, and supersede, our previously published 
measurements~\cite{BaBarAkpiPRL,BaBarPiPi2004}.

We are grateful for the excellent luminosity and machine conditions
provided by our \pep2\ colleagues, 
and for the substantial dedicated effort from
the computing organizations that support \babar.
The collaborating institutions wish to thank 
SLAC for its support and kind hospitality. 
This work is supported by
DOE
and NSF (USA),
NSERC (Canada),
IHEP (China),
CEA and
CNRS-IN2P3
(France),
BMBF and DFG
(Germany),
INFN (Italy),
FOM (The Netherlands),
NFR (Norway),
MIST (Russia),
MEC (Spain), and
PPARC (United Kingdom). 
Individuals have received support from the
Marie Curie EIF (European Union) and
the A.~P.~Sloan Foundation.

%


\begin{thebibliography}{99}
\bibitem{refConj}
The use of charge-conjugate modes is implied throughout this Letter unless otherwise noted.
\end{thebibliography}
\end{document}